\documentclass[preprint]{aastex701}

\graphicspath{{./}{figures/}}

\usepackage{comment}
\usepackage{appendix}

\newcommand{\Lsun}{\mbox{$L_{\odot}$}}

\newcommand{\Lstar}{\mbox{$L_{\ast}$}}

\newcommand{\Teff}{\mbox{$T_{\rm eff}$}}
\newcommand{\teff}{\mbox{$T_{\rm eff}$}}
\newcommand{\logg}{\mbox{$\log$~\textsl{g}}}

\newcommand{\logL}{\mbox{$\log L_{\ast}/L_{\odot}$}}
\newcommand{\Lbol}{\mbox{$L_{\rm bol}$}}

\newcommand{\vinf}{\mbox{$v_{\infty}$}}

\newcommand{\mdot}{$\dot M$}
\newcommand{\Mdot}{$\dot M$}
\newcommand{\vsini}{\mbox{$v \sin i$}}

\newcommand{\kps}{km~s$^{-1}$}

\newcommand{\Myr}{\mbox{$ M_{\odot} \; yr^{-1}$}}

\newcommand{\ha}{$\rm H{\alpha}$}

\newcommand{\kms}{$\rm km\,s^{-1}$}

\newcommand{\ba}{$\rm Br{\alpha}$}

\received{}
\revised{}
\accepted{}
\submitjournal{ApJ}

\shorttitle{JWST observations of SMC OB~stars}
\shortauthors{Garcia et al.}

\begin{document}

\title{Probing thin winds in massive OB stars of the SMC with JWST/NIRSPEC $\mathbf{\rm Br{\alpha}}$\ spectroscopy}

\correspondingauthor{Miriam Garcia}

\author[0000-0003-0316-1208]{Miriam Garcia}
\affil{Centro de Astrobiología, CSIC-INTA;   
Crtra. de Torrejón a Ajalvir km 4; 
28850 Torrejón de Ardoz (Madrid), Spain}
\email{mgg@cab.inta-csic.es}

\author[0000-0003-2429-7964]{Alexander W. Fullerton}
\affil{Space Telescope Science Institute;
3700 San Martin Drive;
Baltimore, MD 21218, USA}
\email{fullerton@stsci.edu}

\author[0000-0002-9124-0039]{Francisco Najarro}
\affil{Centro de Astrobiología, CSIC-INTA;
Crtra. de Torrejón a Ajalvir km 4;
28850 Torrejón de Ardoz (Madrid), Spain}
\email{najarro@cab.inta-csic.es}

\author[0000-0002-0874-1669]{Joachim Puls}
\affil{LMU M\"unchen, Universit\"atssternwarte;
Scheinerstrasse 1;
81679 M\"unchen, Germany}
\email{}

\author[0000-0003-3063-4867]{Daniel J. Lennon}
\affil{Instituto de Astrof\'isica de Canarias;
Avenida V\'ia L\'actea s.n.;
38205 La Laguna, Tenerife, Spain}
\affil{Dpto. Astrof\'isica, Universidad de La Laguna; E-38\,205 La Laguna, Tenerife, Spain}
\email{}

\author[0000-0003-4946-2591]{Jean-Claude Bouret}
\affil{ Aix Marseille Univ, CNRS, CNES, LAM; Marseille, France}
\email{}

\author[0009-0005-7288-6407]{Christopher J. Evans}
\affil{European Space Agency (ESA), ESA Office; Space Telescope Science Institute, 3700 San Martin Drive; 
Baltimore, MD 21218, USA}
\email{}

\author[0000-0003-1780-6150]{Margaret Hanson}
\affil{Department of Physics, University of Cincinnati; P.O. Box 210011; Cincinnati, OH 45221-0011, USA}
\email{}

\author[0000-0001-8768-2179]{Artemio Herrero}
\affil{Dpto. Astrof\'isica, Universidad de La Laguna; E-38\,205 La Laguna, Tenerife, Spain}
\affil{Instituto de Astrof\'isica de Canarias;
Avenida V\'ia L\'actea s.n.;
38205 La Laguna, Tenerife, Spain}
\email{}

\author[0000-0001-5094-8017]{D. John Hillier}
\affil{Department of Physics and Astronomy \& Pittsburgh Particle Physics, Astrophysics, and Cosmology Center (PITT PACC)}
\email{}

\author[0000-0003-2372-9825]{Alexandre Legault}
\affil{Centro de Astrobiología, CSIC-INTA;
Crtra. de Torrejón a Ajalvir km 4;
28850 Torrejón de Ardoz (Madrid), Spain}
\affil{Departamento de F\'isica Te\'orica, Universidad Aut\'onoma de Madrid (UAM);
 Campus de Cantoblanco E-28049 Madrid, Spain}
\email{}


\author[0000-0003-4076-7313]{Mar\'ia del Mar Rubio D\'iez}
\affil{Centro de Astrobiología, CSIC-INTA;
Crtra. de Torrejón a Ajalvir km 4;
28850 Torrejón de Ardoz (Madrid), Spain}
\email{}

\author[0000-0003-1168-3524]{Sergio Sim\'on-D\'iaz}
\affil{Instituto de Astrof\'isica de Canarias;
Avenida V\'ia L\'actea s.n.;
38205 La Laguna, Tenerife, Spain}
\affil{Dpto. Astrof\'isica, Universidad de La Laguna; E-38\,205 La Laguna, Tenerife, Spain}
\email{}

\author[0000-0003-1729-1273]{Jon O. Sundqvist}
\affil{KU Leuven, Instituut voor Sterrenkunde; Celestijnenlaan 200D, 3001 Leuven, Belgium}
\email{}

\author[0000-0001-8631-7700]{Frank Tramper}
\affil{Centro de Astrobiología, CSIC-INTA;
Crtra. de Torrejón a Ajalvir km 4;
28850 Torrejón de Ardoz (Madrid), Spain}
\email{ftramper@cab.inta-csic.es}



\begin{abstract}
Mass loss is a key physical process in the evolution of massive stars, the impact of which propagates into galactic evolution, population synthesis models, the interpretation of high-redshift galaxies, 
and explosive events such as supernovae.
However, there are currently substantial uncertainties in the low-metallicity, 
low-luminosity thin wind regime where classical diagnostics (\ha\ and UV P~Cygni profiles) yield wind momenta that are one to two orders of magnitude below
prescriptions
implemented by
default in most evolutionary models.
Here, we present 
spectra of the mass-loss diagnostic line \ba\ in 15 OB-type stars in the Small Magellanic Cloud obtained using the NIRSpec instrument on the James Webb Space Telescope. 
The line profile morphology, recovered by virtue of the outstanding signal-to-noise ratio of the data and the avoidance of regions with nebular emission, is 
consistent with predictions based on previous mass-loss rate estimates from optical and UV spectroscopy. 
Moreover,   an initial spectroscopic analysis of sources covering the thin wind regime confirms 
  the low mass-loss rates, indicates a change of slope in the wind-momentum luminosity relation 
in this regime with respect to high luminosity objects, and strengthens the above-mentioned discrepancies
with commonly used wind-momentum prescriptions.

%
\end{abstract}




\section{Introduction} 
\label{s:intro}

Mass loss is a key agent in the life-cycle of massive stars \citep[][]{C75,M81} that is widely implemented in stellar evolution codes with simplifying recipes \citep{Lal88,Pal05,Eal08,Pal13}.
However, uncertainties in the dependence of mass-loss rate (\Mdot) on metallicity ($Z$), luminosity (\logL) and effective temperature (\teff), even for stars on the main sequence, have significant implications for their subsequent evolutionary pathways and end products (e.g., \citealt{josiek2024}). 
These \Mdot\ recipes 
rely on
theoretical predictions but are anchored to observations in the Milky Way and the 
  Large and Small Magellanic Clouds (LMC/SMC).
While results for  mass-loss rates and their $Z$-dependence for
high luminosity O-type stars (e.g., \citealt{puls1996,mal07b}) were found to be in reasonable agreement with theoretical predictions \citep{Kudritzki1987}, significant discrepancies   seemed to exist for lower luminosity O-type stars within the Milky Way \citep{Mal04,Marcolino2009} and SMC \citep{Bal03,Ramachandran2019,Rickard2022}, with  estimates of
wind momenta up to two orders of magnitude below 
commonly employed predictions (see below).
This discrepancy in the \emph{thin wind} regime represents a significant observational challenge that, if confirmed by other methods, would have 
far-reaching 
impact on astrophysics in general, 
particularly in the context of feedback from a population of massive stars,
given that later-type, less luminous O-stars are significantly more numerous than higher-mass, earlier-type stars.

\begin{figure}
  \includegraphics[width=0.7\textwidth]{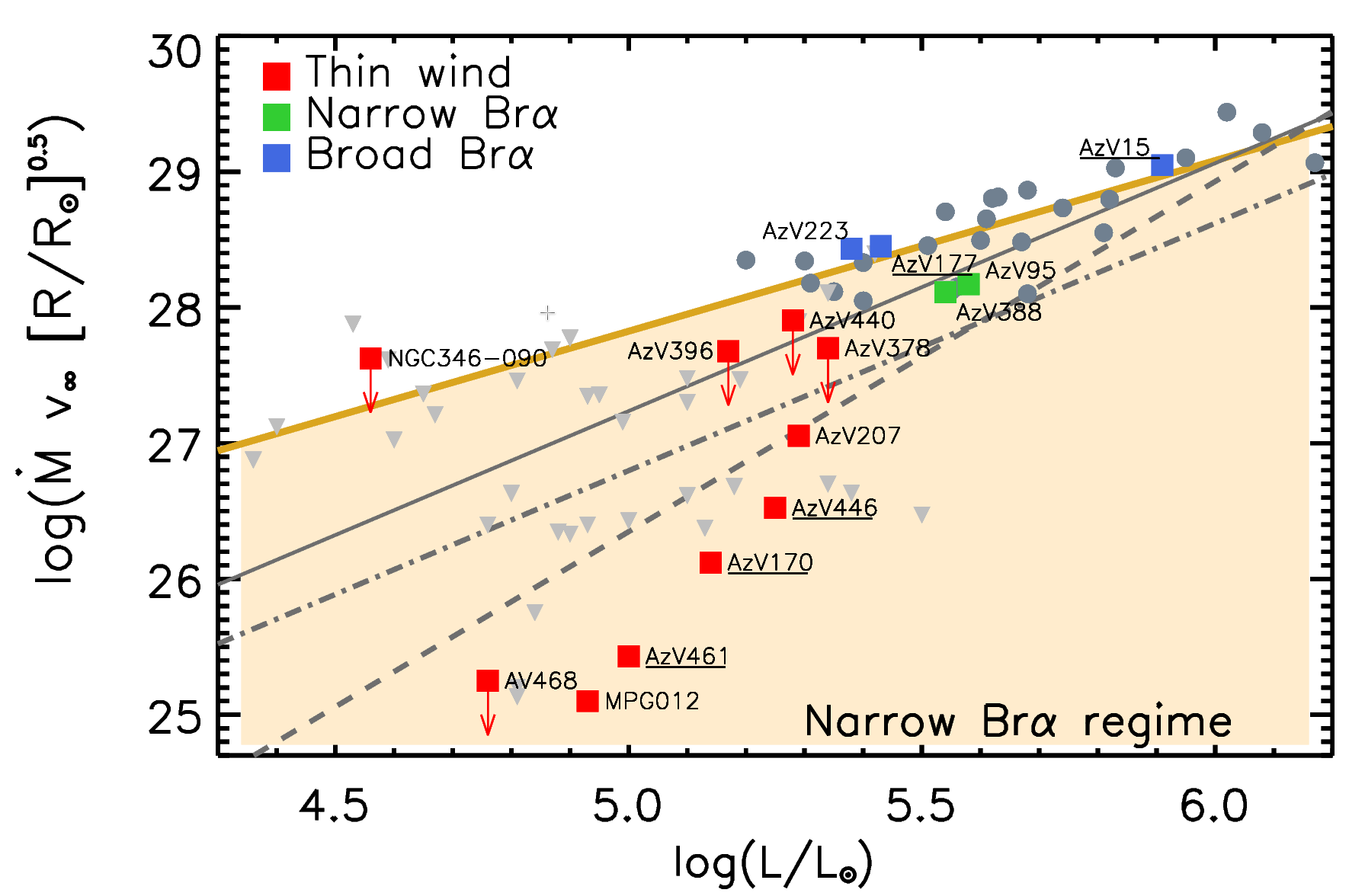}
\caption{
  Wind-momentum luminosity relation at SMC metallicity expressed in \textit{cgs} units.
Gray points mark empirical results, with \Mdot\
obtained from \ha\ or combined optical-UV analyses \citep[][]{Mal06,mal07b,PMal04,PMal05,PMal09,Bal13,HLH06}.
When these works considered clumping, we computed the equivalent smooth-wind 
mass-loss rate
\citep[$\dot M_{smooth}=\dot M/ \sqrt{f_{vol}}$, with $f_{vol}$\ being the clumping volume filling factor, see e.g.][Sect. 6.3]{PulsVinkNajarro2008}
prior to calculating the wind-momentum rate.
\ha\ only provides upper limits for mass loss rates at \logL\, $\lesssim \,$ 5.2--5.4 (downward triangles).
We note the change of slope between the high luminosity objects and the less luminous 
ones at this $L_{\ast}$.
The gray lines represent the theoretical WLR derived by \citet[][solid]{VKL01} and scaled to the SMC metallicity, and
more modern relations at SMC metallicity by \citet[][dash-dotted]{Krticka2018} and \citet[][dashed]{BSPN21}.
The shaded region below the yellow line marks the parameter space where \ba\ is predicted to exhibit narrow emission.
Blue and green symbols denote 
project
targets expected to show broad and narrow \ba\ emission respectively, 
while the red sources lie in the expected thin wind regime (see Table\,\ref{T:targets}
for the stellar parameters used in this plot).
For some project sources, only upper limits of \Mdot\, could
be constrained by previous works (red downward arrows).
Objects with underlined designations have been analyzed in Sect.\,\ref{s:analysis}.  
 }
      \label{fig:wlr}
\end{figure}

The SMC, 
due to its low metallicity ($Z$\,$\sim$\,0.2\,$Z_{\odot}$),
is of special interest as the thin wind regime becomes apparent at earlier spectral types than in the Milky Way,
being present
even for O7 stars on the main sequence with log(\Lstar /\Lsun)\,$\lesssim$\,5.4 \citep[e.g.][]{Marcolino2022,Rickard2022,backs2024}.
This is illustrated in Fig.~\ref{fig:wlr} where we plot the wind-momentum luminosity relation (WLR) for the SMC adopting literature results that are based on optical (\ha), or combined optical-ultraviolet (UV) analyses. 
At \logL $\, \lesssim \,$ 5.2--5.4, only upper \Mdot\ limits can be provided in many cases, most of which are well below
  the widely used \cite{VKL01} relation.
One implication of this discrepancy is that the mass-loss rates of low-luminosity, low-Z massive stars 
could be overestimated by as much as factors of 50--100
when using the prescriptions by \cite{VKL01}, which are 
implemented as default in most models of massive star evolution.
More modern prescriptions 
by \citet[][]{Krticka2018} and \citet{BSPN21} partly reconcile the \Mdot\ differences,
although they do not reproduce the change of slope of the \textit{observational} WLR
at the luminosities when thin winds set in.

Nonetheless, major caveats also affect optical/UV analyses 
that result in a large scatter of the observed \Mdot - \Lbol\
relation at the low luminosity end \citep[see e.g. Fig.\,13 of ][]{BSPN21}.
On the one hand, \ha\ provides only upper limits to \Mdot\ due to its profile being almost purely photospheric in nature, while UV resonance lines exhibit very weak, or entirely absent, P-Cygni features \citep{walborn2000,evans2004} that also depend strongly on both X-ray emission from wind-embedded shocks, and wind inhomogeneities (\citealt{Hillier2003,fullerton2006, Oskinova2007, PulsVinkNajarro2008} and references therein; \citealt{PrinjaMassa2010,SundqvistPulsOwocki2014}).

The Brackett\,$\alpha$ ({\ba}) line at 4.0512 $\rm \mu$m, corresponding to the $n = 5$ -- $4$ transition of atomic hydrogen, offers an alternative and independent method of deriving mass loss rates. 
For dense winds, \ba\ behaves similarly to \ha, i.e., it displays broad
emission, that increases in strength with increasing \Mdot.  
However, for the weaker winds of low luminosity stars, this inherently
strong transition 
exhibits
extreme non-LTE effects particularly in the transition zone between the photosphere and wind.
This results in a strong depopulation of level $n=4$ compared to level $n=5$ and thus an extremely large source function
\citep[appendix~B]{NHP11}. As a consequence, \ba$\!$ exhibits a narrow
emission spike near the line center, while the line wings are
still in absorption. 
\citet{NHP11} showed that the   line-intensity of the central spike of narrow \ba$\!$ profiles increases
as the mass-loss rate decreases, thus providing reliable diagnostics
for extremely low \Mdot\ beyond the sensitivity range of \ha\
\citep[see Fig.\,11 of][]{NHP11},
and that it is relatively insensitive to wind inhomogeneities.
While this is an extremely powerful diagnostic, the thermal background of the Earth’s atmosphere prevents its application 
from ground-based telescopes for all but a handful of targets in the Milky Way \citep{MDA94,NHP11}.

The superb sensitivity of the James Webb Space Telescope (JWST) in the thermal IR
uniquely enables observation of \ba\ in massive stars in the SMC \citep[see also][]{MangRoman2025}.
In this paper we present, for the first time, very high signal-to-noise ratio (S/N$\sim$170--200) observations of 15 OB stars in the SMC taken with the JWST Near Infrared Spectrograph (NIRSpec). 
We discuss whether the \ba\ morphology agrees with 
theoretical predictions from {\sc cmfgen} \citep{HM98} and {\sc fastwind} \citep{FW05, rivero2011} non-LTE, 
  unified atmosphere-plus-wind models, 
and its consistency with the UV and optical diagnostics.
  Building upon the thorough tests on the \ba\ behavior we performed in \citet{NHP11},
we carry out a preliminary analysis of a representative subsample of five sources.
The paper is structured as follows:
the sample, observing configuration, and data reduction are described in Sects.~\ref{s:targets} and \ref{s:datared};
the \ba\ atlas and   analysis 
are presented in Sects.~\ref{s:atlas} and \ref{s:analysis}.
Finally, we provide our conclusions and future work in Sect.~\ref{s:conclusions}.

\section{Sample Selection}
\label{s:targets}

Our primary project goal is to 
obtain   reliable 
\mdot\ measurements of thin wind stars, by exploiting the unique physical behaviour of \ba,
which is
independent from the limitations and uncertainties of \ha\ and UV diagnostics. To this end, the target sample
was designed to probe the mass-loss rates and WLR of O and early B-type stars,
with emphasis on the low luminosity, thin wind regime
(see Table~\ref{T:targets} and Fig.~\ref{fig:wlr}).

The bulk of the sample consists of 
ten stars with \logL$\, \lesssim \,$ 5.4  (labeled `T'), for which
\ha\ is unable to yield mass-loss rates, and where {\sc fastwind} and {\sc cmfgen} models predict narrow \ba\ emission   on top of photospheric 
absorption wings.
In order to enable cross-calibration against
the new \ba\ measurements,
the sample also   includes 
stars with wind-sensitive \ha\ profiles.
  We considered a number of such objects for which
the models predict \ba\ hydrogen emission to be narrow
(`N',   \textit{with} photospheric absorption, e.g. AzV95) or broad 
(`B', \textit{with} both core and wing emission,  e.g. AzV15).
Because some of them are
close to the narrow to broad transition mass loss rates,
and rotational broadening additionally 
affects
the \ba\ profile (see Sect.~\ref{s:atlas}), 
five objects were 
included with various \vsini\ values.
These five targets ensure a proper understanding of the physics of \ba\ and its transition from the narrow to the broad regime.

All targets are well-observed O-type stars in the main body of the SMC, with reliable
photospheric parameters (\Teff, \logg, \logL) derived from optical
or optical and UV data (see Table~\ref{T:targets}). 
Besides luminosity and mass loss rates, the sample was designed to assess the role of effective temperature and gravity.
Binaries were avoided, at least in so far as this information was available. 
A final, critical constraint was that nebulosity should be avoided, which is especially important to prevent nebular contamination of narrow \ba\ profile components. 

\begin{deluxetable*}{l l c c c c c c l c c c r c}
\tablecaption{
  Sample stars sorted by spectral type and luminosity class, with $K_{s}$\ and \textit{Spitzer} 4.5$\mu$m magnitudes from \citet{Bonanos2010}.
  Stellar (\vsini, \Teff, centrifugally corrected \logg$_{true}$, \logL) and wind (\Mdot, \vinf) 
  parameters were adopted from the references provided in column ``Ref.".
  These used \ha\ and/or UV resonance lines as the main mass-loss diagnostics, as follows,
  and considered clumping unless otherwise noted:
  (1) \citet{PMal04}, \ha\ (unclumped); (2) \citet{PMal05}, \ha\ (unclumped); (3) \citet{Mal06}, \ha\ (unclumped); (4) \citet{Bal13}, UV;
  (5) \citet{PMal13}, \ha\ ;   (6) \citet{Bal21}, UV; (7) \citet{Mal24}, UV; (8) \citet{backs2024}, UV+\ha.
  Column ``Regime" specifies the expected wind regime and \ba\ morphology: 
   B -- broad, N -- narrow, and   T -- thin wind (see Sect.~\ref{s:targets}).
  A subsample of five objects was analysed in Sect.~\ref{s:analysis} with {\sc cmfgen} and {\sc fastwind}, 
  yielding the mass-loss rates 
  provided in the 2nd and 3rd entry of column ``\mdot", respectively.
\label{T:targets}}
\tablewidth{700pt}
\tabletypesize{\scriptsize}
\tabcolsep1mm
\tablehead{
\colhead{Target} & \colhead{Spectral Type} & \colhead{$K_{s}$} & 
\colhead{$[4.5]$} & \colhead{$v$sin$i$} &
\colhead{\Teff} & \colhead{log $g_{true}$} & \colhead{log $L_{\ast}/L_{\odot}$} & \colhead{\Mdot}  & 
\colhead{$v_{\infty}$} & \colhead{Regime} & \colhead{Ref.} \\
\colhead{} & \colhead{}& \colhead{$[mag.]$} & 
\colhead{$[mag.]$} & \colhead{$[km$ $s^{-1}]$} &
\colhead{$[K]$} & \colhead{$[dex]$} & \colhead{$[dex]$} & \colhead{[\Myr]}  & 
\colhead{$[km$ $s^{-1}]$} & \colhead{} & \colhead{} 
}
\startdata
AzV 177     & O4V~((f))  & 15.32 & 15.24 & 220 & 44500 & 4.03 & 5.43 & 1.4E-07 \textit{5.0E-08 2.0E-07}& 2400 & B & 4    \\  
AzV 388     & O4~V       & 14.80 & 14.80 & 150 & 43100 & 4.01 & 5.54 & 1.0E-07                         & 2100 & N & 4    \\  
AzV 15      & O6.5~III(f)& 13.61 & 13.60 & 110 & 39750 & 3.70 & 5.91 & 6.8E-07 \textit{2.9E-07 4.5E-07}& 2250 & B & 8    \\ 
AzV 446     & O6.5~V     & 15.42 & 15.37 & 30  & 39700 & 4.00 & 5.25 & 4.0E-09 \textit{3.0E-09 3.0E-09}& 1400 & T & 4    \\ 
AzV 207     & O7~V((f))z & 14.92 & 15.05 & 110 & 38000 & 3.82 & 5.29 & 2.2E-08                         & 1775 & T & 8    \\ 
AzV 95      & O7.5~V((f))& 14.25 & 14.20 & 75  & 38250 & 3.64 & 5.58 & 7.6E-08                         & 1850 & N & 8    \\  
AzV 461     & O8~V       & 15.41 & 15.56 & 200 & 37100 & 4.05 & 5.00 & 1.0E-09 \textit{1.3E-09 1.3E-09}& 1540 & T & 4    \\
AzV 440     & O8~V       & 15.27 & 15.30 & 100 & 37000 & 4.01 & 5.28 & $\la$\, 3.0E-07                 & 1300 & T & 2    \\ 
AzV 468     & O8.5~V     & 15.92 & 15.95 & 50  & 34700 & 4.00 & 4.76 & $\la$\, 7.0E-10                 & 1540 & T & 4    \\ 
AzV 223     & O9.5~II    & 14.30 & 14.25 & 140 & 31600 & 3.45 & 5.37 & 2.0E-07                         & 1680 & B & 5    \\  
AzV 378     & O9.5~III   & 14.48 & 14.53 & 110 & 31500 & 3.27 & 5.34 & $\la$\, 1.0E-07                 & 2000 & T & 1, 2 \\
NGC346-ELS-090
            & O9.5~V     & 16.48 & 16.43 & 188 & 34900 & 4.28 & 4.56 & $\la$\, 9.8E-08                 & 2978 & T & 3    \\ 
AzV 170     & O9.7~III   & 14.67 & 14.71 & 70  & 30500 & 3.41 & 5.14 & 1.5E-09 \textit{2.5E-09 2.5E-09}& 1200 & T & 6    \\ 
AzV 396     & B0~III     & 15.04 & 14.92 & 120 & 30000 & 3.52 & 5.17 & $\la$\, 1.0E-07                 & 2000 & T & 1, 2 \\ 
NGC346-MPG012
           & B0~IV      & 15.35 & 14.79 &  60 & 31000 & 3.65 & 4.93 & 5.0E-10                         & 1250 & T & 4    \\
\enddata
\end{deluxetable*}


\begin{deluxetable*}{llclcccrc}
\tablecaption{Journal of Observations. See main text for column description.\label{logfile}}
\tablehead{\colhead{Target}         & 
           \colhead{Spectral Type}  & 
           \colhead{VISIT ID}       & 
           \colhead{READPATT}       & 
           \colhead{NINTS}          & 
           \colhead{NGROUPS}        & 
           \colhead{UT(START)}      & 
           \colhead{EXPTIME [s]}    &  
           \colhead{S/N}
          }
\startdata
AzV 177        & O4 V((f))   & 03225012001 & NRSIRS2      & 2 & 16 & 2024-06-25T21:11:26 & 11671 & 184 \\
AzV 388        & O4 V        & 03225013001 & NRSIRS2RAPID & 1 & 90 & 2024-08-03T14:25:11 &  6565 & 181 \\
AzV 15         & O6.5 III(f) & 03225015001 & NRSIRS2RAPID & 1 & 30 & 2024-06-25T17:43:52 &  2188 & 170 \\
AzV 446        & O6.5 V      & 03225008001 & NRSIRS2      & 2 & 18 & 2024-08-03T06:17:55 & 13130 & 184 \\
AzV 207        & O7 V((f))z  & 03225009001 & NRSIRS2      & 1 & 20 & 2024-07-31T19:42:33 &  7294 & 175 \\
AzV 95         & O7.5 V((f)) & 03225014001 & NRSIRS2      & 1 & 12 & 2024-08-23T02:39:05 &  4377 & 166 \\
AzV 461        & O8 V        & 03225004001 & NRSIRS2      & 2 & 18 & 2024-07-04T20:15:12 & 13130 & 185 \\
AzV 440        & O8 V        & 03225007001 & NRSIRS2      & 2 & 16 & 2024-08-03T02:23:44 & 11671 & 186 \\
AzV 468        & O8.5 V      & 03225002001 & NRSIRS2      & 3 & 20 & 2024-07-04T13:28:23 & 21883 & 189 \\
AzV 223        & O9.5 II     & 03225011001 & NRSIRS2RAPID & 1 & 60 & 2024-08-03T12:37:56 &  4377 & 188 \\
AzV 378        & O9.5 III    & 03225010001 & NRSIRS2RAPID & 1 & 70 & 2024-08-03T10:36:02 &  5106 & 186 \\
NGC346-ELS-090 & O9.5 V      & 03225001001 & NRSIRS2      & 6 & 20 & 2024-08-02T10:16:31 & 43767 & 213 \\
AzV 170        & O9.7 III    & 03225005001 & NRSIRS2RAPID & 1 & 80 & 2024-06-25T18:55:36 &  5836 & 180 \\
AzV 396        & B0 III      & 03225006001 & NRSIRS2      & 2 & 13 & 2024-08-02T23:10:13 &  9483 & 188 \\
NGC346-MPG-012 & B0 IV       & 03225003001 & NRSIRS2      & 2 & 16 & 2024-07-31T15:53:50 & 11671 & 181 \\
\enddata
\end{deluxetable*}

\section{Observations and Data Reduction}
\label{s:datared}

Table~\ref{logfile} summarizes the JWST/NIRSpec data obtained under 
the auspices of JWST Cycle 2 program GO-03225 (PI: Garcia).
All observations were made in Fixed Slit mode through the S200A2 aperture with 
the high-resolution G395H grating and the F290LP long-pass filter.
This configuration provides wavelength coverage between 2.87 and 5.27~$\rm \mu$m with 
a small gap between 3.81 and 3.91~$\rm \mu$m due to the separation of the two detectors.
Besides \ba, our observations include additional diagnostic lines such as Pf${\gamma}$\,3.75$\rm \mu$m, Hu${\beta}$\,4.65$\rm \mu$m, He{\sc ii}\,3.09\,$\rm \mu$m, He{\sc i}\,4.29\,$\rm \mu$m and He{\sc ii}\,4.76\,$\rm \mu$m.
On-orbit measurements indicate that the resolving power of this mode is $\sim$3600 
\citep{Isobe2023}
which is considerably better than pre-launch predictions of $\sim$2700 
\citep{Jakobsen2022}.
This resolving power corresponds to velocity resolution of $\sim$85~{\kms} at the wavelength of
{\ba}.   

Each observational sequence consisted of a wide-aperture target acquisition (WATA) to place
the source accurately in the S200A2 aperture, which was followed by spectroscopic exposures
at 5 primary dither  positions offset along the spatial dimension of the aperture.
The NIRSpec detectors were read with either the NRSIRS2RAPID or NRSIRS2 pattern to mitigate the effects of “1/f noise”.  
NRSIRS2RAPID was generally preferred for integrations consisting of fewer than 90 groups. 
Table~\ref{logfile} records the number of integrations (NINTS) and groups per integration (NGROUPS) 
that were tuned to achieve 
S/N\,$\geq$\,150
without saturating, 
as required to define the continuum flux to a precision of 1\%.
The total exposure time devoted to each target and the median S/N per pixel 
measured between 4.2 and 4.4 $\rm \mu$m are listed in the final two columns of Table~\ref{logfile}.

The observations were processed with version 1.18.0 of the JWST calibration pipeline,
which uses the reference files specified in Calibration Reference Data System (CRDS) 
context jwst\_1364.pmap.
Processing through the initial {\tt CALWEBB\_DETECTOR1} stage relied on default values of 
parameters to produce a count-rate image for each exposure in the dither pattern
for a given target, and included additional mitigation for 1/f noise in the
{\tt clean\_flicker} step.
These images were subsequently flat-fielded, rectified and resampled in 
the {\tt CALWEBB\_SPEC2} stage of the pipeline.
A combined background image was created and subtracted before the individual images were
combined and a single spectrum extracted during processing in   {\tt CALWEBB\_SPEC3} stage.
Default values were generally chosen for all these processing stages, which used ``exptime'' 
weighting in both the {\tt resample\_spec} and {\tt outlier\_detection} steps.
However, the extraction aperture was manually recentered on the trace in the 
resampled and rectified ``s2d'' image in the {\tt CALWEBB\_SPEC3} processing.

\begin{figure}
\includegraphics[width=0.45\textwidth]{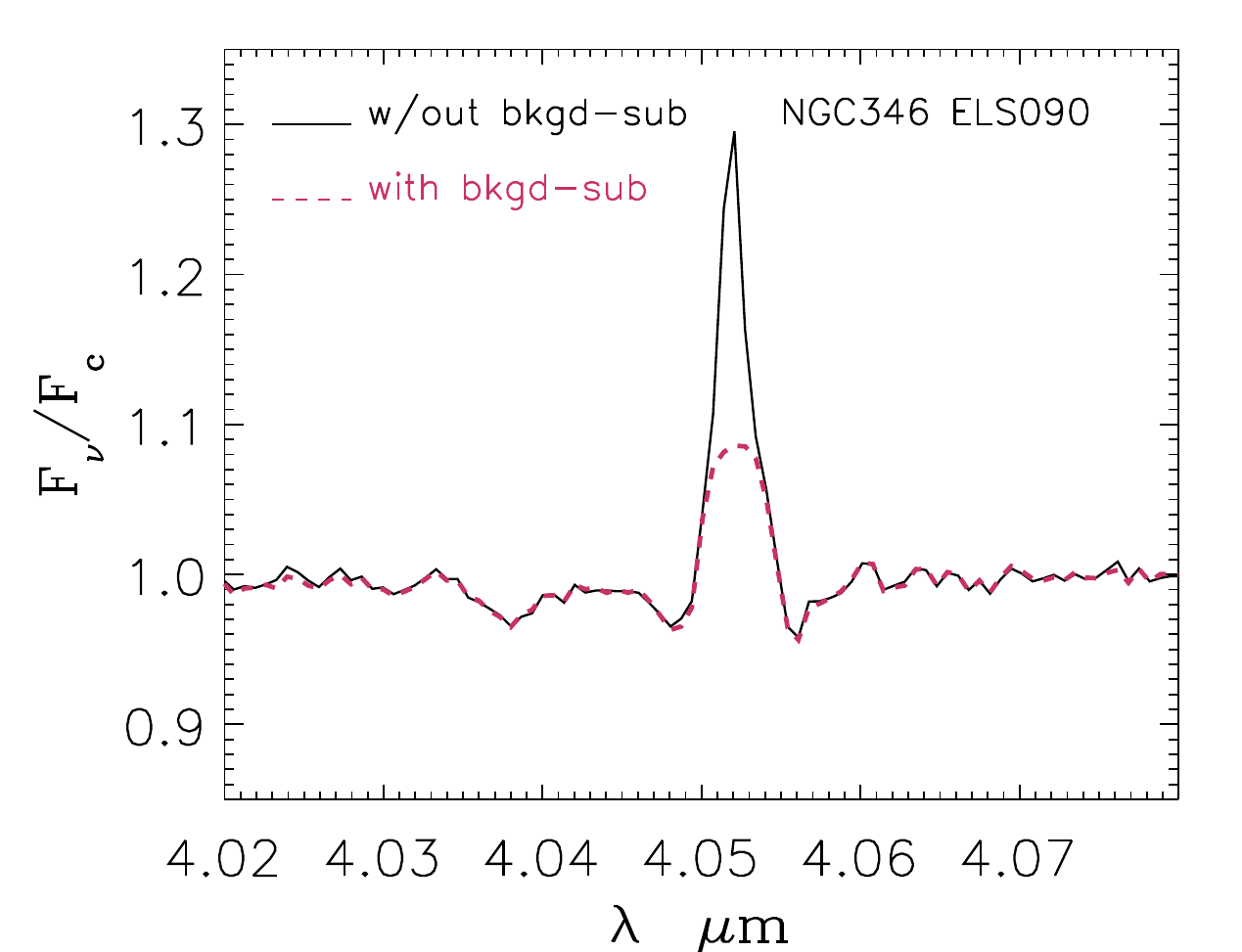} 
\includegraphics[width=0.4\textwidth]{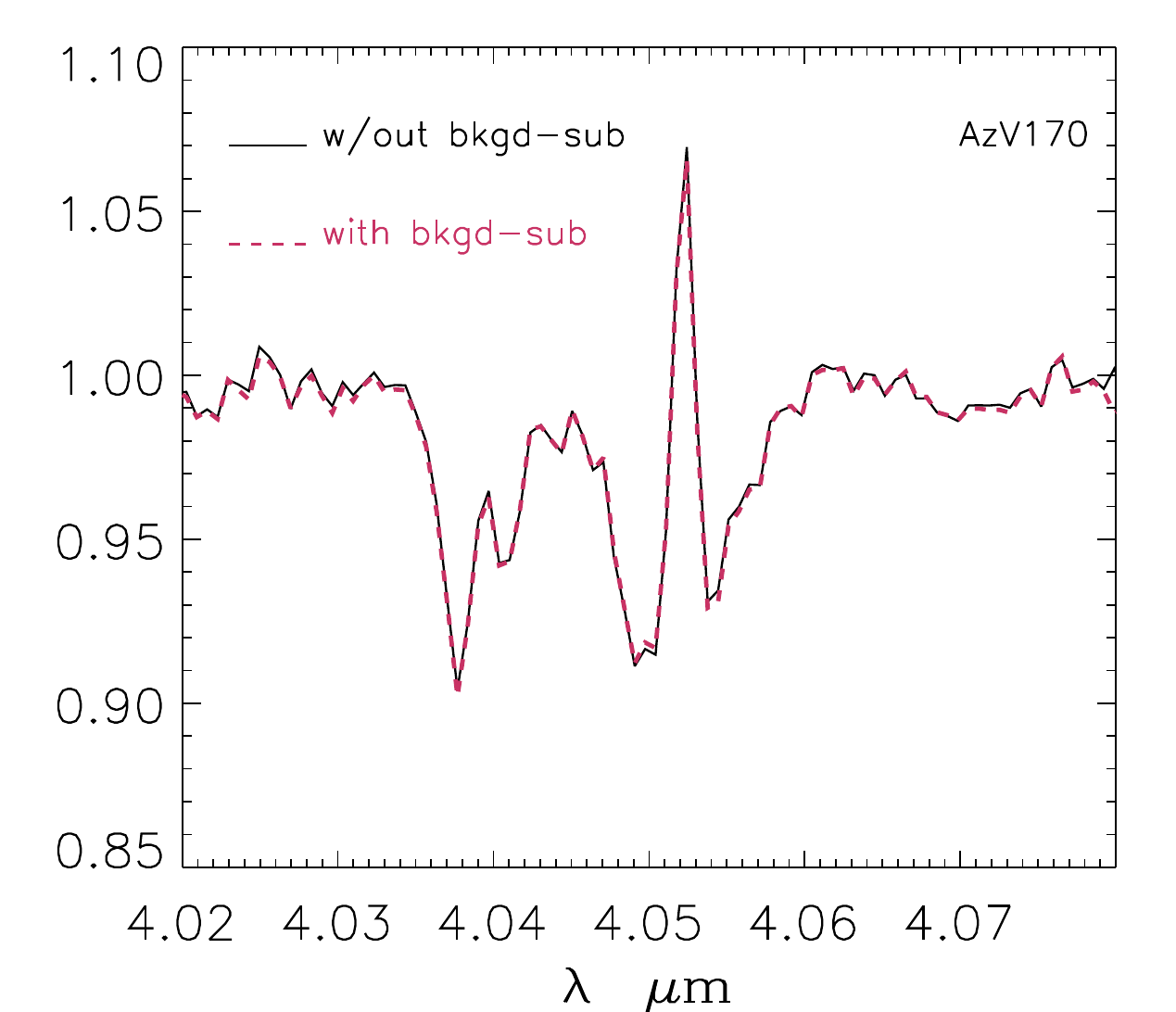} 
\caption{Left: NIRSpec spectrum of NGC346-ELS-090, embedded in intense H{\sc ii} emission, 
               after reduction with and without background subtraction. 
          Right: Same, now showing a target within negligible nebulosity.
      \label{F:bkgd}}
\end{figure}

Our background subtraction strategy proved to be a critical step in recovering the shape of the stellar {\ba}\ profile.
Figure~\ref{F:bkgd} illustrates the influence of nebular contamination on \ba\
for a target on the outskirts of the giant H~{\sc ii} region N66 (NGC346-ELS-090; left panel) and for a more isolated target (AzV170; right panel).
In both cases the strategy of constructing a mean background image from the 5 spatially dithered images 
provided adequate mitigation of astrophysical contamination. 
This procedure, together with the consideration of the nebular environment during target selection,
allowed the difficulties posed by background contamination described by \citet{MangRoman2025} to be avoided.

\section{\ba$\!$~morphology of SMC O-type stars}
\label{s:atlas}

\begin{figure}
\includegraphics[width=0.49\textwidth]{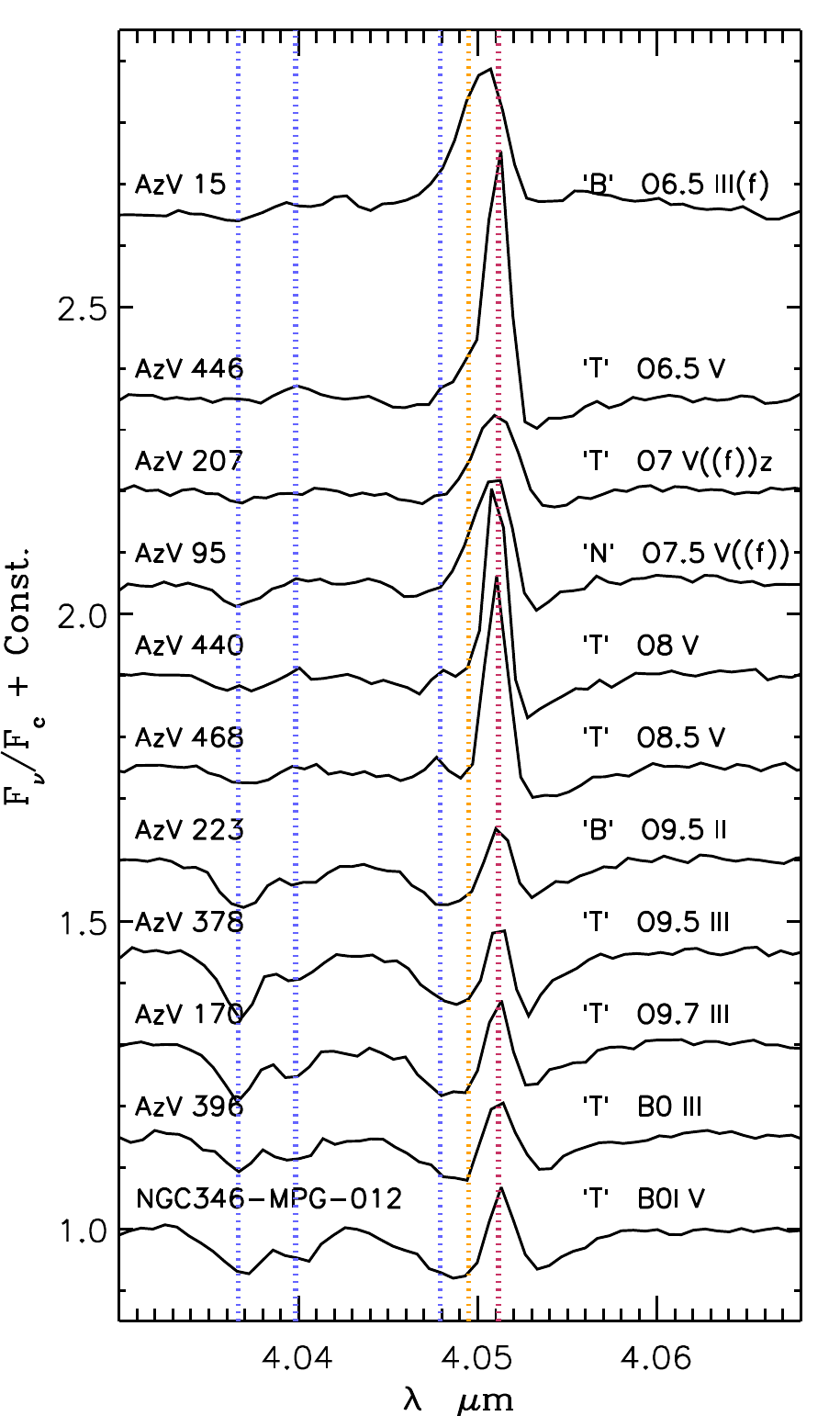}
\caption{\ba\ 
spectra of slowly rotating sample stars, arranged by spectral type and luminosity class.
The spectra are shown in the same vertical scale, offset by a constant for clarity.
'B', 'N' and 'T' specify the \textit{expected} 
regime.
Vertical lines mark H{\sc i} (red), He{\sc i} (blue) and He{\sc ii} (orange)  transitions. 
  Stark absorption wings underlying the \ba\ emission
are absent in AzV15, and then evolve 
towards strong absorption in the later types.
}
      \label{F:atlas1}
\end{figure}

\begin{figure}
\includegraphics[width=0.49\textwidth]{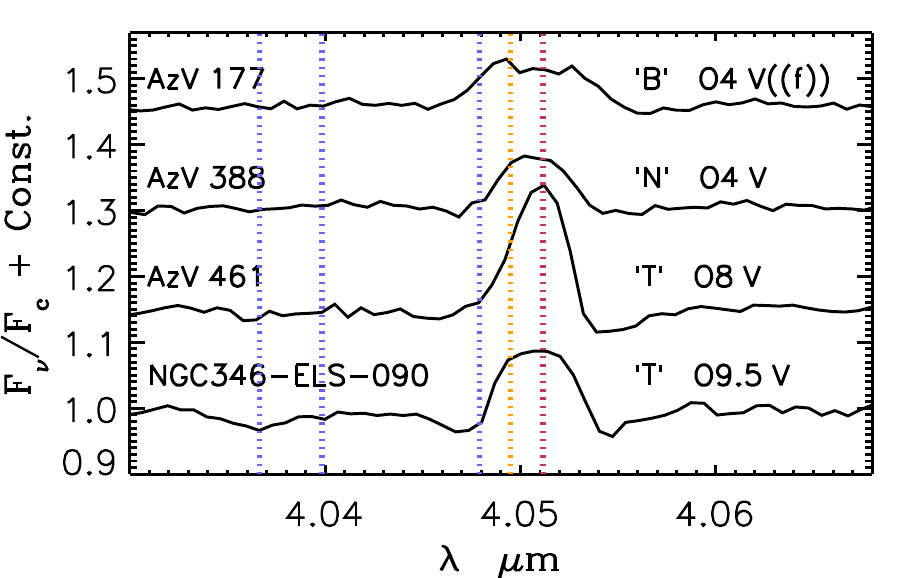}
\caption{Same as Fig.~\ref{F:atlas1}, now  
for the sample stars with \vsini $\geq$ 150 \kms. 
The high rotational velocities broaden the \ba\ profiles and cause the \textit{apparent}
disagreement between observations and the prediction. 
The late spectral types show the broad absorption wings seen in Fig.~\ref{F:atlas1},
  although somehow washed out by the fast rotation.
      \label{F:atlas2}}
\end{figure}

Figure~\ref{F:atlas1} displays the NIRSpec \ba\ observations of the sample stars with relatively low projected rotational speed \vsini,
arranged by spectral type and luminosity class.
Because rotational velocity broadens the \ba\ emission, the observations of fast rotating,
\vsini $\geq$ 150 \kms\
stars are shown separately in Fig.~\ref{F:atlas2}.

Figure~\ref{F:atlas1} shows 
the trend of the line morphology with spectral type, though somewhat 
distorted 
by rotation and macroturbulence.
 AzV15 exhibits pure, broad \ba\ emission.
For the rest of the stars,
\ba\ shows core emission on top of broad Stark absorption wings
that 
increase in strength towards later spectral subtypes.
%

For the hotter types, the overlapping He{\sc ii}\,4.0495\,$\rm \mu$m emission leads   to broader emission (e.g., AzV95 and AzV207) and to a  shift to
shorter
wavelengths of the peak of the \ba\  line complex
(e.g. AzV15 ) 
as well as to an overall asymmetry in both broad, and pure-narrow emission profiles.
For the later types, the asymmetry in the absorption component is caused by the He{\sc i}\,4.048\,$\rm \mu$m $5f$ -- $4f$ triplet and singlet transitions (note also the strong He{\sc i} $n = 5$ -- $4$ hydrogenic transitions at 4.037, and 4.040~$\rm \mu$m in these subtypes). Thus, whereas the blue side of the \ba\ line complex is typically contaminated by He{\sc ii} (close to line center) and/or He{\sc i} (in the absorption wings), its red side  is dominated by the hydrogen transition alone, leading to the asymmetric line shape.



The observed morphology matches the model predictions of narrow/broad \ba\ profiles as listed in Table~\ref{T:targets},
except for AzV223.
Broad \ba\ emission was expected for this star, but it exhibits
 the narrow emission and broad photospheric absorption typical for late types.
This object displays emission filling in the \ha\ absorption core, so that 
an actual \Mdot\ value 
could be derived under the assumption that the emission
originated in the wind  (see Table~\ref{T:targets}).
However, the observed \ba\ line strongly hints at lower \Mdot\ than expected from \ha.

\section{\Mdot\ constraints from quantitative spectroscopy}
\label{s:analysis}

\begin{figure}
\includegraphics[width=0.49\textwidth]{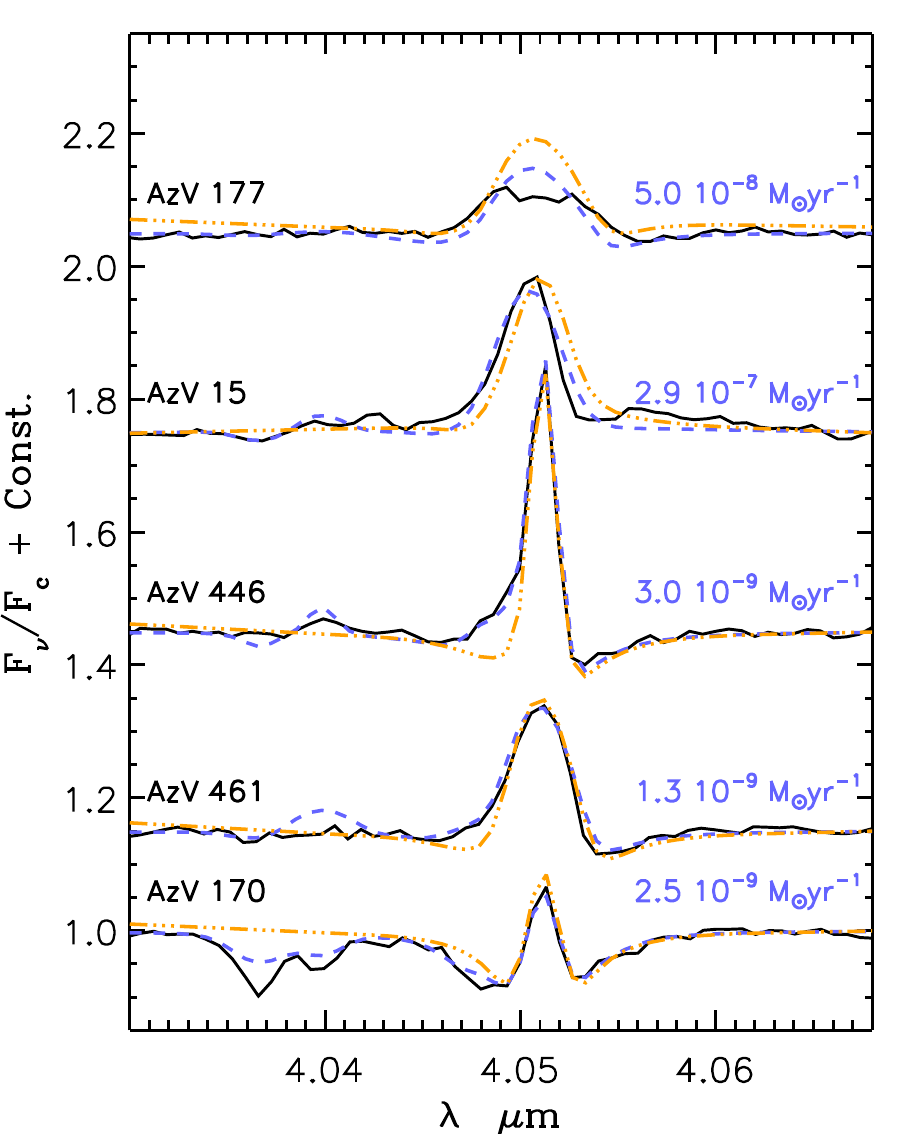} 
\caption{
  Line fits to a subset of five sample stars, encompassing different wind regimes,
with {\sc cmfgen} (blue) and {\sc fastwind} (orange) models. 
The mass loss rate values obtained by the analysis with {\sc cmfgen} are provided next to each star.
      \label{F:preliminary}}
\end{figure}

We performed an initial analysis for a subsample of five stars with 
previously determined
UV mass-loss rates,   spanning different temperatures, luminosities and
wind regimes, 
by means of the non-LTE, unified model atmosphere code {\sc cmfgen}. 
  Our panchromatic analysis by visual comparison against
UV, optical and JWST/NIRSpec data resulted 
in roughly the same stellar parameters as provided in the literature (see Table~\ref{T:targets}).
To obtain an overall good fit 
we adapted the wind parameters, 
the transition velocity between photosphere and wind, the microturbulence, and the clumping law (the latter to fit optical and UV wind lines),
aiming at the first 
atmosphere models that reproduce the complete UV to IR spectrum including the critical \mdot\ indicator \ba.
  An example fit is shown in Figure\,\ref{fig:fit}.
  Pathfinder models 
show that neither clumping nor X-ray emission from wind-embedded shocks affect \ba\ in the thin wind regime, since it is formed in the transonic region where inhomogeneities and shocks are weak or absent  (see also Fig.\,\ref{fig:fit}).
They further show that the impact of X-rays on \ba\ is negligible in stronger winds as well.
  More details will be provided in a forthcoming paper but we refer the reader to \citet{NHP11} for thorough tests on the impact of clumping and transition velocity.

As an independent check, we used {\sc fastwind} (v10.6) models to calculate \ba\ profiles, 
for identical photospheric parameters, microturbulence, and clumping law as derived by {\sc cmfgen}, and varied only \Mdot\ and the shape of the velocity law (keeping {\sc fastwind}'s default transition velocity at $v_{\rm trans}=$ 0.1 $v_{\rm sound}$).

  Fig.~\ref{F:preliminary} shows the best fitting models for the five stars analyzed at the \ba\ region.
After accounting for \vsini, macroturbulence and spectral resolution, 
all {\sc cmfgen} and most {\sc fastwind} profiles reproduce 
the observed \ba\ emission and (if present) 
the asymmetric broad photospheric absorption 
quite well.
  Despite extremely thin winds, 
we managed to constrain mass-loss rates with both codes and obtained similar or identical values (see Table~\ref{T:targets}).
The most outstanding difference, of a factor 4, is registered for AzV177,
but this fast rotator could experience spherical-symmetry departures.
In fact, the shape of the \ba\ emission resembles several spectral features
observed in the optical spectrum of the fast rotating O4~If $\zeta$~Puppis star \citep{BouretHillier2012}.
Moreover, we currently estimate  \Mdot\ uncertainties as roughly a factor 2, given the influence
of the transition velocity and the microturbulence on the formation of \ba\ in thin wind stars   \citep{NHP11}.
Further tests and accurate error estimates will be performed in the forthcoming paper.

Most interestingly, however, our \Mdot\ values generally agree well with those from previous studies using UV lines (see Table~\ref{T:targets}). Thus it seems that at least for 
this subset of five
objects, the uncertainties discussed in Sect.~\ref{s:intro} do not play an important role, and/or the underlying assumptions for UV analyses (X-rays, clumping) 
are reasonable.
Moreover, our results confirm the low \Mdot\ derived for thin wind stars in the UV,
and the discrepancy with 
\citet{VKL01}'s WLR
by $\sim$1--2 orders of magnitude.

\section{Conclusions and future work}
\label{s:conclusions}
We present high S/N JWST/NIRSpec observations of a sample of thin wind candidates in the SMC. 
In contrast to \citet{MangRoman2025}, we find a clear correlation of \ba\ morphology
with spectral types, likely emerging as a result of
the higher S/N 
and minimal nebular contamination issues of our dataset.
Moreover, a comparison between the observed morphology and predictions 
from model atmosphere codes reveals 
good   
agreement.

For a subsample of five objects, we analyzed archival UV and optical spectra together with our new IR data, 
to infer the mass-loss rate   mainly 
from \ba.
The resulting \Mdot\ values are quite similar to previous values obtained from the UV alone.
Overarching conclusions from our analysis are: i) \ba\ is a  useful 
\Mdot\ estimator also in the case of thin winds;
ii) narrow \ba\ can provide mass-loss rates in the range where other diagnostics are insensitive 
or not available 
(e.g., due to high extinction);
and iii) there is overlap with the regime where \ha\ \textit{is} sensitive, thus enabling a critical consistency check
between mass-loss rates derived from narrow \ba\ and \ha.

One of the major questions to be answered is whether
thin winds are actually weak, or only mimic such low \Mdot\ because of insufficient shock-cooling and/or 
the presence of multiple components \citep[e.g.][]{Huene2012, Bouret2015, lagae2021}.
In this context \citet{Law2024} used JWST/MIRI to observe highly ionized, forbidden mid-infrared emission lines that form in the \textit{outer} wind of 10\,Lac, and obtained a mass-loss rate one order of magnitude above UV/optical results.
While very promising, this method is subject to uncertainties related to clumping and multiple wind components, which likely have little impact on the \ba\ diagnostics that are based on the conditions in the transonic region.
Nevertheless, additional data in this spectral domain would be informative.

Analysis of our complete dataset, including the full list of lines formed in the transonic region
and captured by our JWST/NIRSpec data (see Sec.~\ref{s:datared}),
will provide unique insights on the true mass loss rates of thin wind stars
and potentially yield a robust, empirical WLR over the entire luminosity range.
Furthermore, comparison with Milky Way analogues will enable a new determination of the metallicity dependence of radiation-driven mass-loss, providing the boundary conditions for empirically supported mass-loss recipes in codes simulating the evolution of massive stars in the nearby and distant Universe.

Given the somewhat surprising result that all five
analysed stars display consistent \ba\ and UV mass-loss rates, 
and that the observational features for the other objects also point to very low mass-loss rates for $\log(L/L_\odot) \le 5.4$,
it is highly likely that the weak-wind problem is real:
at low luminosities and low metallicities,
the values expected from an extrapolation of the empirical high luminosity WLR 
are too large, and 
  modern models fail to reproduce the observed change of slope,
requiring further effort for clarification. 
In this sense, the problem remains unsolved.

\section*{Acknowledgments}
This work is based on observations with the NASA/ESA/CSA James Webb Space Telescope obtained at the Space Telescope Science Institute, which is operated by the Association of Universities for Research in Astronomy, Incorporated, under NASA contract NAS5-03127. 
All data products can be accessed at MAST: \dataset[10.17909/xxf8-f637]{http://dx.doi.org/10.17909/xxf8-f637}.
Support for Program number 03225 is provided through a grant from the STScI under NASA contract NAS5-03127.

M. Garcia, F. Najarro, A. Legault and F. Tramper gratefully acknowledge support by grant PID2022-137779OB-C41, and M. Garcia further acknowledges grant PID2022-140483NB-C22, 
funded by the Spanish Ministry of Science, Innovation and Universities/State Agency of Research 
MICIU/AEI/10.13039/501100011033 and by “ERDF A way of making Europe”.  M. Garcia and F. Najarro also acknowledge grant MAD4SPACE, TEC-2024/TEC-182 from Comunidad de Madrid (Spain). 
D.J. Lennon, A. Herrero and S. Sim\'on-D\'iaz are supported by the Spanish Government Ministerio de Ciencia e Innovaci\'on through grant PID2021-122397NB-C21. 
A. Legault also acknowledges funding from grant PREP2022-000263.
A. W. Fullerton and D.J. Hillier acknowledge support from grants JWST-GO-03225.002-A and JWST-GO-03225.003-A, respectively.
  We gratefully acknowledge valuable comments and suggestions by our anonymous referee.

\facility{JWST (NIRSpec)}

\bibliography{ms_v0_MAIN.bib}
\bibliographystyle{aasjournal}

\appendix
 
\section{Multiwavelength Fit}
\begin{figure*}
\includegraphics[width=0.9\textwidth]{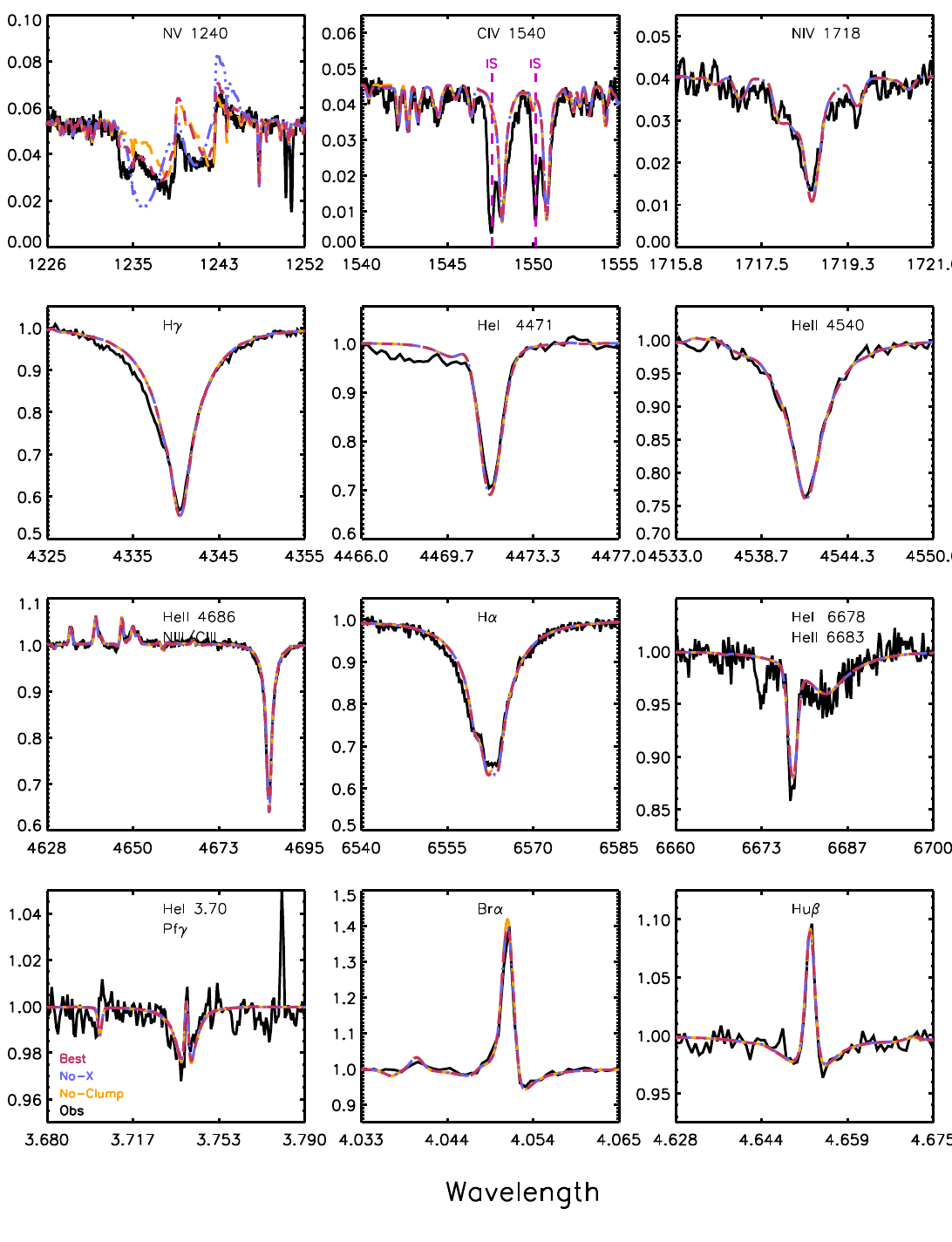} 
\caption{  JWST observations of AzV446 and archival optical and UV data (black),
together with model fits that account for instrumental and rotational/macroturbulence broadening.
The best-fitting CMFGEN model (red) reproduces \Teff\ and \logg\ diagnostic lines
in the visible, as well as classical mass-loss rate indicators in the UV and optical ranges. 
The model spectra successfully reproduce the observed \mdot-sensitive lines in the L-band (lowermost row).
For comparison, the figure
includes synthetic spectra for models without clumping (orange) and without X-rays (blue). Such differences in the simulations do not produce any effect on the displayed L-band lines.
 }
      \label{fig:fit}
\end{figure*}

To illustrate the 
quality of the analysis described in Sect.\,\ref{s:analysis}, 
Fig.\,\ref{fig:fit} shows the best fitting synthetic spectra to AzV446
for important stellar and wind diagnostic lines observed in the 
UV \citep[ULLYSES library,][]{Roman-Duval2025}, optical \citep[XSHOOTU library,][]{Vink2023} and L-band ranges (this work).
The fit includes
the \mdot-sensitive diagnostic
lines Pf${\gamma}$, \ba\ and Hu${\beta}$.
The best fitting {\sc CMFGEN} model has very similar parameters to those found by \citet{Bal13}:
slightly adjusted \teff=40750~K and \logg=4.2,
the same clumping and velocity field values these authors derived, $f_{vol}$=0.1, \vinf=1400\kps,
and X-ray luminosity $L_x/L_{\ast}$ =5E-07, $v_{\rm trans}=$ 0.1 $v_{\rm sound}$,
and \Mdot=3.0E-09~\Myr.
Additional
{\sc CMFGEN} spectra for models with the same parameters, except for clumping and X-rays which are switched off, are  also shown in Fig.\,\ref{fig:fit} 
to show that related variations of \ba\ (and other L-band lines) are negligible.

\end{document}